\documentclass[preprint,12pt]{elsarticle}  
\usepackage{graphicx} 
\usepackage{amssymb} 
\usepackage{amsmath} 
\usepackage{amsfonts} 
\usepackage{slashed}
\usepackage[dvipsnames]{xcolor} 
\newcommand{\nopieft}{\mbox{$\slashed{\pi}$EFT}~}

\journal{Physics Letters B} 

\begin{document} 

\begin{frontmatter} 

\title{Onset of $\eta$-nuclear binding in a pionless EFT approach} 
\author[a]{N.~Barnea} 
\author[b]{B.~Bazak} 
\author[a]{E.~Friedman} 
\author[a]{A.~Gal\corref{cor1}} 
\address[a]{Racah Institute of Physics, The Hebrew University, 91904 
Jerusalem, Israel}
\address[b]{IPNO, CNRS/IN2P3, Univ. Paris-Sud, Universit\'{e} 
Paris-Saclay, F-91406 Orsay, France} 
\cortext[cor1]{corresponding author: Avraham Gal, avragal@savion.huji.ac.il} 

\begin{abstract} 
$\eta NNN$ and $\eta NNNN$ bound states are explored in stochastic variational 
method (SVM) calculations within a pionless effective field theory (EFT) 
approach at leading order. The theoretical input consists of regulated $NN$ 
and $NNN$ contact terms, a regulated {\it energy dependent} $\eta N$ contact 
term derived from coupled-channel models of the $N^{\ast}(1535)$ nucleon 
resonance plus a regulated $\eta NN$ contact term. A self consistency 
procedure is applied to deal with the energy dependence of the $\eta N$ 
subthreshold input, resulting in a weak dependence of the calculated 
$\eta$-nuclear binding energies on the EFT regulator. It is found, in terms 
of the $\eta N$ scattering length $a_{\eta N}$, that the onset of binding 
$\eta\,^3$He requires a minimal value of Re$\,a_{\eta N}$ close to 1~fm, 
yielding then a few MeV $\eta$ binding in $\eta\,^4$He. The onset of binding 
$\eta\,^4$He requires a lower value of Re$\,a_{\eta N}$, but exceeding 0.7~fm. 
\end{abstract} 


\end{frontmatter}

\section{Introduction} 
\label{sec:intro} 

The $\eta N$ $s$-wave interaction near threshold, $E_{\rm th}(\eta N)=
1487$~MeV, is attractive as realized first by coupling to the $\pi N$ 
channel~\cite{BLi85} and subsequently confirmed, e.g.~\cite{WKW96}, by 
coupling the $\eta N$ channel to the entire set of meson-baryon channels 
with $0^-$ octet mesons and $\frac{1}{2}^+$ octet baryons, thereby generating 
dynamically the $N^{\ast}$(1535) $S_{11}$ resonance. The size of the resulting 
$\eta N$ energy dependent $s$-wave attraction, however, is strongly model 
dependent with values of the real part of the $\eta N$ scattering length as 
low as 0.2~fm~\cite{WKW96} and up to nearly 1~fm in the $K$-matrix model of 
Green and Wycech (GW)~\cite{GWy05} and in the recent Giessen coupled channels 
study~\cite{SLM13}. Following the work of Ref.~\cite{BLi85} it was soon 
realized that $\eta$-nuclear quasibound states might exist~\cite{HLi86,LHa86} 
with widths determined by the scale of the imaginary part of the $\eta N$ 
scattering length. This imaginary part, due mostly to $\eta N\to \pi N$, 
is small in all models, between 0.2 to 0.3~fm. Nevertheless, no $\eta$-nuclear 
quasibound state has ever been established beyond doubt~\cite{Wilkin16}. 

Recent optical-model calculations of such bound states~\cite{FGM13,CFG14} 
using several energy-dependent $\eta N$ model amplitudes are summarized in 
Refs.~\cite{Gal14,Mares16}. Whereas the appearance of $\eta$-nuclear 
bound states is robust in any of these $\eta N$ interaction models, 
the value of mass number $A$ at which binding begins is model dependent. 
Thus, the relatively strong $\eta N$ attraction in model GW even admits in 
such calculations a $1s_{\eta}$ bound state in $^4$He, with as low binding 
energy as 1.2~MeV and width of 2.3~MeV~\cite{Mares16} calculated using 
a static $^4$He density. Unfortunately, the $\eta$-nucleus optical model 
approach is not trustable for as light nuclei as $^4$He, and genuine 
few-body calculations are required. 

Photon- and hadron-induced reactions on nuclear targets provide useful 
constraints on possible $\eta$ bound states in very light nuclei, where 
according to a recent review~\cite{KWi15} the most straightforward 
interpretation of the data is that $\eta d$ is unbound, $\eta\,^3$He is nearly 
or just bound, and $\eta\,^4$He is bound. Our previous few-body $\eta NN$ and 
$\eta NNN$ calculations~\cite{BFG15}, using the Minnesota~\cite{MNC77} and 
Argonne AV4'~\cite{AV402} $NN$ potentials, agree with this conjecture as far 
as the $\eta d$ and $\eta\,^3$He systems are concerned. A similar conclusion 
for $\eta\,^3$He has been reached recently by evaluating the $pd\to \eta\,^3
$He near-threshold reaction~\cite{XLO16}. And a recent WASA-at-COSY search for 
a possible $\eta\,^4$He bound state in the $dd\to\,{^3{\rm He}}N\pi$ reaction 
placed upper limits of a few nb on the production of a near-threshold bound 
state~\cite{AAB16}. On the theoretical side, no precise few-body calculation 
of $\eta NNNN$ bound-state has ever been reported for $\eta\,^4$He.{\footnote
{A very recent preprint by Fix and Kolesnikov \cite{Fix17} reports on 
few-body calculations of the $\eta\,^3$He and $\eta\,^4$He scattering 
lengths, concluding that these systems are unbound.}} 

The present work reports for the first time on precise few-body $\eta NNNN$ 
calculations in which the Stochastic Variational Method (SVM) is applied to 
$\eta$ plus few-nucleon Hamiltonians constructed in Leading Order (LO) within 
a Pionless Effective Field Theory. While this \nopieft approach has been 
applied before to few-nucleon systems, e.g.~\cite{Kol99,BHK00}, and more 
recently in lattice-nuclei calculations~\cite{BCG15,KBG15}, it is extended 
here to include constituent pseudoscalar mesons for which pion exchange with 
nucleons is forbidden by parity conservation of the strong interactions. In 
particular, the single $\eta N$ contact term required in LO is provided 
by the $\eta N$ $s$-wave scattering amplitude $F_{\eta N}(E_{\rm sc})$ at 
a subthreshold energy $E_{\rm sc}$ derived self consistently within the 
few-body calculation, as practised in our previous work~\cite{BFG15}. This 
is demonstrated in two $\eta N$ interaction models, GW~\cite{GWy05} and CS 
(Ciepl\'{y}--Smejkal \cite{CSm13}), which exhibit strong energy dependence of 
$F_{\eta N}(E)$ arising from the proximity of the $N^{\ast}$(1535) resonance. 
The results of the few-body calculations reported in the present work suggest 
that the onset of binding $\eta\,^3$He requires a value of Re$\,a_{\eta N}$ 
close to 1~fm, whereas the onset of binding $\eta\,^4$He requires a somewhat 
weaker $\eta N$ interaction with Re$\,a_{\eta N}$ exceeding 0.7~fm.

\section{Methodology} 
\label{sec:meth} 

Here we outline the methodology of the present work, including the few-body 
SVM used, the choice of $NN$, $NNN$, $\eta N$ and $\eta NN$ \nopieft regulated 
contact terms, and the self consistent treatment of the energy-dependent 
$\eta N$ term.

\subsection{SVM calculations} 
\label{sec:SVM} 

SVM calculations were introduced in the mid seventies to few-body nuclear 
problems~\cite{KKr77}, and used extensively with correlated Gaussian bases 
since the mid 1990s~\cite{VSu95}. The SVM was benchmarked together with six 
other few-body methods in calculating the $^4$He binding energy~\cite{KNG01}. 
Correlated Gaussian trial wavefunctions in this method are written as 
\begin{equation} 
\Psi = \sum_k c_k {\cal A} \left (\left[{\cal Y}_{L}^{k}(\hat{\bf x})
        \chi_{S}^k\right]_{JM} \xi_{T T_z}^k
        {\exp(-{\frac{1}{2}}{\bf x}^T A_k{\bf x})} \right ) 
\label{eq:Psi} 
\end{equation} 
where the summation index $k$ runs with linear variational parameters $c_k$ 
on all possible values of the total spin $S$ and the total orbital angular 
momentum $L$, as well as on all possible intermediate coupling schemes, 
$\chi_S$ and $\xi_T$ stand for spin and isospin functions of the $N$-particle 
system, respectively, ${\cal Y}_L$ is the orbital part of $\Psi$ formed by 
coupling successively spherical harmonics in the $(N-1)$ relative coordinates 
of which the vector ${\bf x}$ is made, and $\cal A$ antisymmetrizes over 
nucleons. The matrix $A_k$ introduces $N(N-1)/2$ nonlinear variational 
parameters which are chosen stochastically. For a comprehensive review see 
Ref.~\cite{SVa98}.

\subsection{Pionless EFT nuclear interactions} 
\label{sec:NN} 

Here we follow a \nopieft approach at LO. To this order the nuclear 
interaction consists of two-body and three-body contact (zero-range) terms, 
\begin{equation} 
V_2(ij)=\left[c^\Lambda_S\,\frac{1}{4}(1-{\bf\sigma}_i\cdot{\bf\sigma}_j) + 
c^\Lambda_T\,\frac{1}{4}(3+{\bf\sigma}_i\cdot{\bf\sigma}_j)\right]\,
\delta_{\Lambda}(r_{ij}),  
\label{eq:ij} 
\end{equation} 
\begin{equation} 
V_3(ijk)=d_{NNN}^\Lambda\,\delta_{\Lambda}(r_{ij},r_{ik}),  
\label{eq:ijk} 
\end{equation} 
where these contact terms are smeared by using normalized-to-one Gaussians 
with a regulating momentum-space scale (cut-off) parameter $\Lambda$: 
\begin{equation} 
\delta_{\Lambda}(r_{ij}) = \left(\frac{\Lambda}{2\sqrt{\pi}}\right)^3\,
\exp \left(-{\frac{\Lambda^2}{4}}r_{ij}^2\right), \,\,\, \delta_{\Lambda}
(r_{ij},r_{ik}) = \delta_{\Lambda}(r_{ij}) \delta_{\Lambda}(r_{ik}), 
\label{eq:rij}
\end{equation}
with $\delta_{\Lambda}(r_{ij})$ in the zero-range limit $\Lambda \to \infty$ 
becoming a Dirac $\delta({\bf r}_{ij})$ function. For a given value of the 
scale parameter $\Lambda$, two-body low-energy constants (LEC) $c^\Lambda_S$ 
and $c^\Lambda_T$ are fitted to the $S=0$ $pn$ scattering length and to 
the $S=1$ deuteron binding energy $B(d)$, respectively, and a three-body 
LEC $d_{NNN}^\Lambda$ is fitted to $B(^3$H). Following Ref.~\cite{KPD17} 
a small corrective proton-proton contact term, with LEC $c^\Lambda_{pp}$, 
is introduced together with the Coulomb interaction between protons to 
reproduce the $^3$He binding energy. These two-body and three-body LECs are 
listed in Ref.~\cite{KPD17} where $c^\Lambda_{pp}$ was found to effectively 
adjust $c^\Lambda_S$ by less than 0.1\% over the full $\Lambda$-range 
tested in the present work. The $^4$He calculated binding energy $B(^4$He) 
provides then a check on how reasonable this LO \nopieft version is. This 
is demonstrated in Table~\ref{tab:B4} for four representative values of 
$\Lambda$. The calculated values of $B(^4$He) depend only moderately on the 
scale parameter $\Lambda$, exhibiting renormalization scale invariance by 
approaching in the limit $\Lambda\to\infty$ a finite value 27.8$\pm$0.2~MeV 
which compares well with $B_{\rm exp}(^4$He)=28.3~MeV, despite the fact that 
only LO contributions are accounted for in this \nopieft version. 

\begin{table}[!h] 
\begin{center} 
\caption{$^4$He binding energies $B(^4$He) (in MeV) in LO \nopieft SVM 
calculations, with LECs fitted to $NN$ and $NNN$ low-energy data~\cite{KPD17}. 
The $\Lambda\to\infty$ limit of $B(^4$He) was evaluated by using higher 
values of the scale $\Lambda$ than listed here.} 
\begin{tabular}{ccccccc} 
\hline\hline 
$\Lambda$~(fm$^{-1})$ & 2 & 4 & 6 & 8 & $\to \infty$ & exp. \\ 
\hline 
$B(^4$He) & 22.4 & 22.9 & 24.2 & 25.1 & 27.8$\pm$0.2 & 28.3 \\ 
\hline\hline
\end{tabular} 
\label{tab:B4} 
\end{center} 
\end{table}

\subsection{Pionless EFT $\eta N$ interactions} 
\label{sec:etaN} 

\begin{figure}[htb]
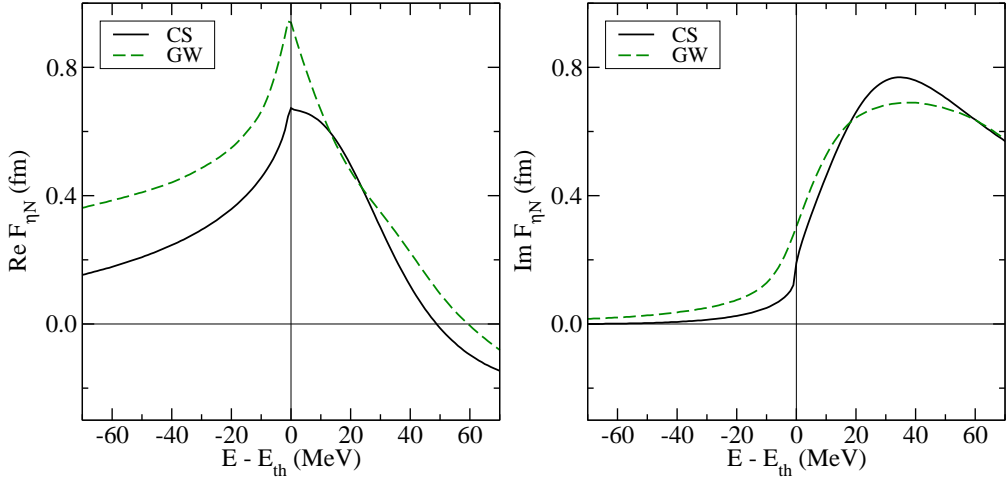
 
\begin{center} 
\includegraphics[width=0.48\textwidth]{retamod2} 
\includegraphics[width=0.48\textwidth]{ietamod2} 
\caption{Real (left panel) and imaginary (right panel) parts of the 
$\eta N$ cm $s$-wave scattering amplitude $F_{\eta N}(E)$ as a function 
of $E-E_{\rm th}$, with $E=\sqrt{s}$ the total $\eta N$ cm energy, in the 
GW~\cite{GWy05} and CS~\cite{CSm13} meson-baryon coupled-channel interaction 
models. The vertical line marks the $\eta N$ threshold energy $E_{\rm th}$. 
Figure adapted from Ref.~\cite{Mares16}.} 
\label{fig:FetaN} 
\end{center} 
\end{figure}

Parity conservation forbids pion exchange in the $\eta N$ interaction, 
suggesting thereby that a \nopieft approach may be justified. With spin and 
isospin zero for the $\eta$ meson, a single two-body $\eta N$ contact term 
is needed at LO. Below we derive the corresponding LEC from the $\eta N$ 
$s$-wave scattering amplitude $F_{\eta N}(E)$ calculated in two meson-baryon 
coupled-channel interaction models, GW~\cite{GWy05} and CS~\cite{CSm13}, 
and shown in Fig.~\ref{fig:FetaN}. Whereas the GW model used in our previous 
work~\cite{BFG15} is an on-shell $K$-matrix model that considers $\eta N
\leftrightarrow\pi N$ coupling, the CS model is a meson-baryon multi-channel 
chirally motivated model in which the $\eta N$ interaction is extremely short 
ranged and practically momentum independent below the momentum breakdown scale 
specified in the next paragraph. Both models capture the main features of the 
underlying $N^{\ast}(1535)$ resonance which peaks about 50~MeV above the $\eta 
N$ threshold energy $E_{\rm th}=1487$~MeV and generates considerable energy 
dependence of $F_{\eta N}$ near threshold. In particular, both Re$\,F_{\eta N}
(E)$ and Im$\,F_{\eta N}(E)$, which at threshold are given by the scattering 
lengths (in fm) 
\begin{equation} 
a_{\eta N}^{\rm GW}=0.96+i0.26 \,\,\,\,\,\, \,\,\,\,\,\, 
a_{\eta N}^{\rm CS}=0.67+i0.20, 
\label{eq:a} 
\end{equation} 
decrease monotonically in these models upon going into the 
subthreshold region while displaying considerable model dependence. 

The $\eta N$ scattering lengths listed above are of order 1~fm or less, 
much smaller than the $NN$ scattering lengths whose large size justifies the 
use of \nopieft in light nuclei. For $\eta N$ interactions as weak as implied 
by this size of $a_{\eta N}$, and with no $\eta N$ bound or virtual state 
expected, the $\eta N$ scattering length alone does not provide a meaningful 
criterion of fitting into an EFT approach. Alternatively, we estimate 
$p_{\eta}R\lesssim\frac{\pi}{2}$ for the momentum $p_{\eta}$ of a weakly 
bound $\eta$-nuclear state in a square well of radius $R$. With $R=2$~fm 
or a bit larger for the He isotopes, we get $p_{\eta}\lesssim 150$~MeV/c 
($\approx$0.76~fm$^{-1}$). Since the lowest-mass allowed meson exchange in 
pseudoscalar meson interaction with octet baryons is owing to vector mesons, 
with a typical mass of $m_{\rho}=770$~MeV, this range of $\eta$-nuclear 
momenta can be accommodated comfortably within the nuclear LO \nopieft 
approach of the preceding subsection. The \nopieft small parameter associated 
with this momentum breakdown scale of $Q^{\rho}_{\rm high}\approx m_{\rho}= 
3.9$~fm$^{-1}$ is given by $(p_{\eta}/Q^{\rho}_{\rm high})^2\approx 0.04$. 

As in previous work \cite{BFG15}, and in order to account for the energy 
dependence inherent in the meson-baryon coupled channel dynamical generation 
of the $N^{\ast}(1535)$ resonance, we construct energy-dependent local 
potentials $v_{\eta N}(E)$ that produce the $\eta N$ energy dependent 
scattering amplitude $F_{\eta N}(E)$ below threshold in models GW and CS. 
For a given $\eta N$ interaction model, the on-shell scattering amplitude 
$F_{\eta N}(E)$ serves as a single datum to which LO \nopieft two-body 
regulated contact terms of the form 
\begin{equation}
v_{\eta N}(E;r)=c^{\Lambda}_{\eta N}(E)\,\delta_{\Lambda}(r), \;\;\;\;\; 
c^{\Lambda}_{\eta N}(E)=-\frac{4\pi}{2\mu_{\eta N}}\,b^{\Lambda}_{\eta N}(E), 
\label{eq:v(E)}
\end{equation}
are fitted. Here $\delta_\Lambda$ is a regulating normalized-to-one Gaussian 
with scale parameter $\Lambda$, as per Eq.~(\ref{eq:rij}), and $c^{\Lambda}_{
\eta N}(E)$ is an energy dependent LEC conveniently related through the $\eta 
N$ reduced mass $\mu_{\eta N}$ to a strength function $b^{\Lambda}_{\eta N}(E)
$ of length dimension. The specific value $c^{\Lambda}_{\eta N}(E_{\rm sc})$ 
of this LEC for a given cut-off $\Lambda$ is determined self consistently 
in the $\eta$-nuclear SVM calculation as detailed in the next subsection. 
By using the {\it same} value of $\Lambda$ in all $NN$, $NNN$, $\eta N$ and 
$\eta NN$ regulating Gaussians we reach a consistent extension of the nuclear 
\nopieft to a combined $\eta$-nuclear \nopieft approach.{\footnote{Along with 
the $NNN$ LEC that averts a Thomas collapse of the $NNN$ system, an $\eta NN$ 
LEC $d_{\eta NN}^{\Lambda}$ is needed to avert $\eta NN$ collapse. Given no 
$\eta NN$ datum, a rough estimate $d_{\eta NN}^\Lambda = d_{NNN}^{\Lambda}$ 
is made here; see Appendix A: Erratum to the PLB published version.}} 

In order to study the renormalization scale invariance of our few-body 
$\eta$-nuclear results, as shown for the purely nuclear case of $B(^4$He) in 
Table~\ref{tab:B4}, we discuss below \nopieft calculations done for several 
representative values of the scale parameter, $\Lambda=2,4,6,8$~fm$^{-1}$. 
The last two values, clearly, exceed the momentum breakdown scale $Q^{\rho}_{
\rm high}\approx 3.9$~fm$^{-1}$ of the underlying $N^{\ast}(1535)$ resonance 
model for the $\eta N$ interaction, or even more so the lower momentum 
breakdown scale $q^{\rho}_{\rm high}\approx 3.0$~fm$^{-1}$ set by excitation 
of vector meson degrees of freedom absent in the underlying $N^{\ast}(1535)$ 
dynamical models considered here, such as the $\rho$ meson produced at 
threshold in the strong pion exchange reaction $\pi N\to\rho N$ with 
$p^{\rho}_{\rm th}=586$~MeV/c. Finally, the model dependence of the LO 
$\eta N$ contact term introduced by studying two quite different $\eta N$ 
interaction models, GW and CS, leaves little motivation to go at present 
beyond LO. Hence, discussion of higher orders in \nopieft is left to future 
work. 

\begin{figure}[!ht]
\begin{center}
\includegraphics[width=0.46\textwidth]{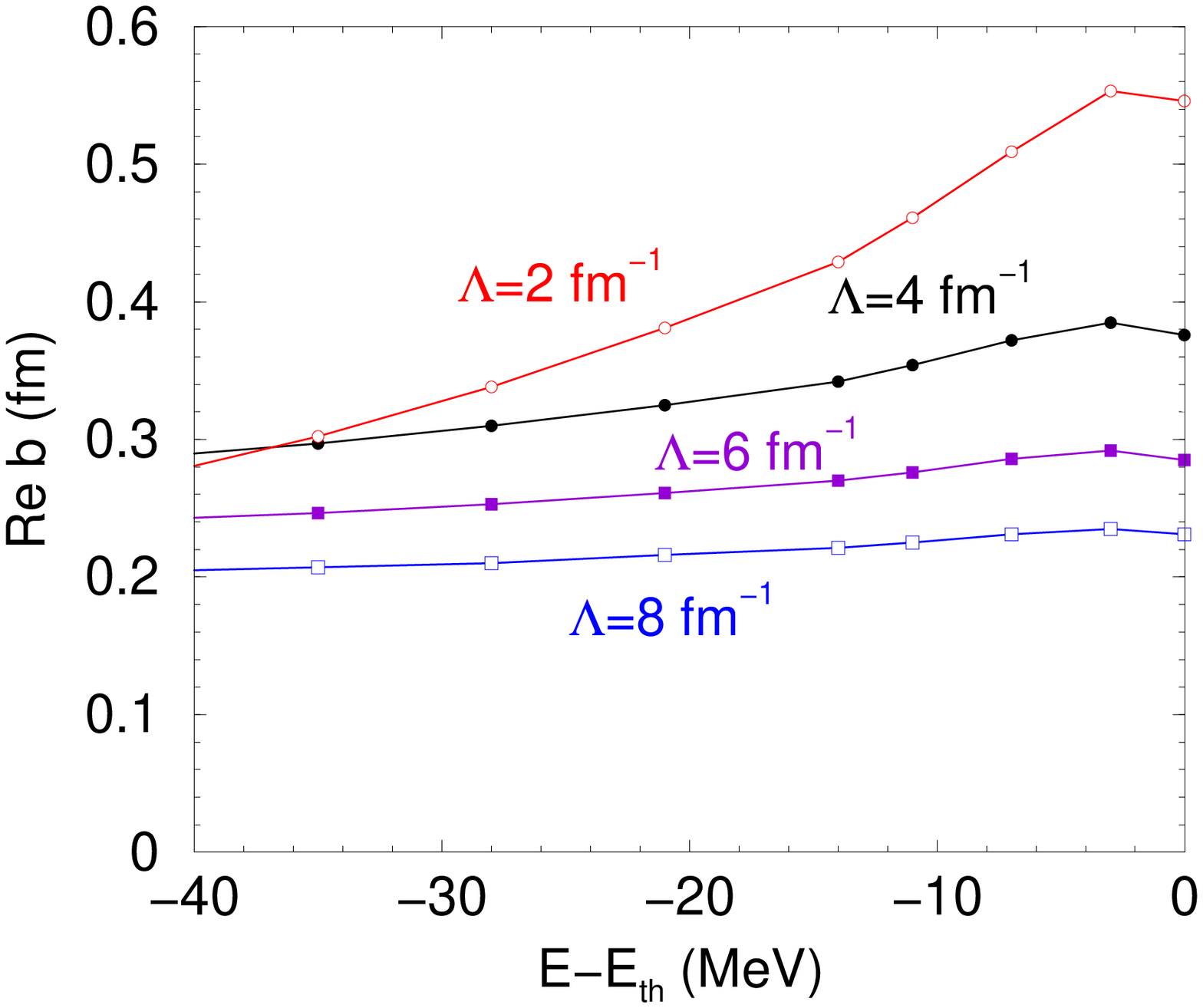}
\includegraphics[width=0.48\textwidth]{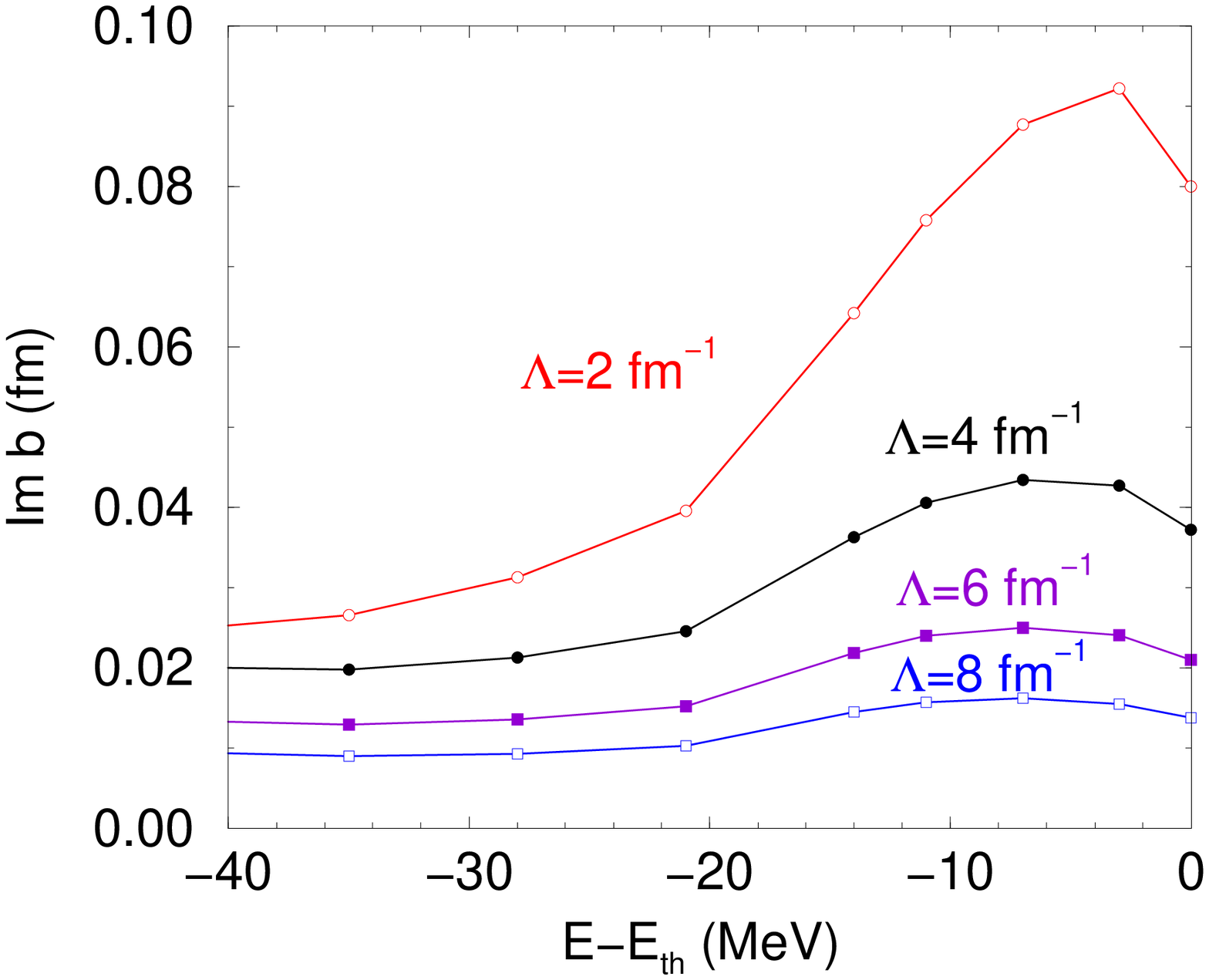}
\caption{Real (left panel) and imaginary (right panel) parts of the strength 
parameter $b^{\Lambda}_{\eta N}(E)$ of the $\eta N$ effective potential 
(\ref{eq:v(E)}) at subthreshold energies $E<E_{\rm th}$ for four values 
of the scale (cut-off) parameter $\Lambda$, all of which result in the 
same scattering amplitude $F_{\eta N}^{\rm GW}$~\cite{GWy05} shown in 
Fig.~\ref{fig:FetaN}.} 
\label{fig:Wycfit8} 
\end{center} 
\end{figure} 

For a given value of $\Lambda$, the subthreshold values of the complex 
strength parameter $b^{\Lambda}_{\eta N}(E)$ in Eq.~(\ref{eq:v(E)}) were 
fitted to the complex phase shifts derived from the subthreshold scattering 
amplitudes $F_{\eta N}(E)$ in models GW and CS. The resulting values of 
the strength parameter $b^{\Lambda}_{\eta N}(E)$ for $\eta N$ subthreshold 
energies in model GW, shown in Fig.~\ref{fig:Wycfit8}, fall off monotonically 
for both real and imaginary parts in going deeper below threshold, 
except for small kinks near threshold that reflect the threshold cusp of 
Re~$F_{\eta N}(E)$ at $E_{\rm th}$ in Fig.~\ref{fig:FetaN}. Similar curves 
for $b^{\Lambda}_{\eta N}(E)$ are obtained in model CS, with values 
smaller uniformly for both real and imaginary parts than model GW yields, 
in accordance with the relative strength of the generating scattering 
amplitudes $F_{\eta N}(E)$ shown in Fig.~\ref{fig:FetaN}. We note that 
increasing $\Lambda$ leads to weaker strengths $b^{\Lambda}_{\eta N}(E)$ and 
also to a weaker energy dependence. Inspecting Fig.~\ref{fig:Wycfit8}, one 
also notes that Im~$b^{\Lambda}_{\eta N}(E)$$\ll$Re~$b^{\Lambda}_{\eta N}(E)$, 
which justifies treating Im~$v_{\eta N}$ perturbatively in the present 
calculations.

\subsection{Energy dependence} 
\label{sec:E} 

To determine the $\eta N$ subthreshold energy at which $v_{\eta N}(E)$ is 
evaluated as input to the $\eta$-nuclear few-body calculations reported 
below, we denote the shift away from threshold by $\delta\sqrt{s}\equiv
\sqrt{s}-\sqrt{s_{\rm th}}$, expressing it in terms of output expectation 
values~\cite{BFG15}:  
\begin{equation} 
\langle\delta\sqrt{s}\rangle = -\frac{B}{A} -\beta_{N}\frac{1}{A}\langle T_N 
\rangle +\frac{A-1}{A}E_{\eta} -\xi_A\beta_{\eta}\left ( \frac{A-1}{A} 
\right )^2 \langle T_{\eta} \rangle \; , 
\label{eq:sqrt{s}} 
\end{equation} 
where $\beta_{N(\eta)}\equiv m_{N(\eta)}/(m_N+m_{\eta})$, $\xi_A\equiv Am_N/
(Am_N+m_{\eta})$, $T_N$ and $T_{\eta}$ are the nuclear and $\eta$ kinetic 
energy operators evaluated in terms of internal Jacobi coordinates, with 
$T=T_N+T_{\eta}$ the total intrinsic kinetic energy of the system, $B$ is 
the total binding energy of the $\eta$-nuclear few-body system and $E_{\eta}=
\langle\Psi|(H-H_N)|\Psi\rangle$, where $H_N$ is the Hamiltonian of the purely 
nuclear part in its own cm frame, and the total Hamiltonian $H$ is evaluated 
in the overall cm frame. The imaginary, absorptive part of the $\eta N$ 
interaction is suppressed in this discussion. Noting that $(A-1)\langle 
T_{N:N}\rangle$ in Eq.~(7) of Ref.~\cite{BFG15} equals $\langle T_N\rangle$ 
here, Eq.~(\ref{eq:sqrt{s}}) coincides with the former equation apart from 
a kinematical factor $\xi_A$ introduced here to make correspondence with the 
$\eta$-nuclear, last Jacobi coordinate with which $T_{\eta}$ is associated. 
Requiring that the expectation value $\langle\delta\sqrt{s}\rangle$ on 
the l.h.s. of Eq.~(\ref{eq:sqrt{s}}), as derived from the solution of 
the Schroedinger equation, agrees with the input value $\delta\sqrt{s}$ 
for $v_{\eta N}(E)$, this equation defines a self-consistency cycle in our 
few-body $\eta$-nuclear calculations. Since each one of the four terms on 
the r.h.s. of Eq.~(\ref{eq:sqrt{s}}) is negative, the self consistent energy 
shift $\delta\sqrt{s_{\rm sc}}$ is also negative, with size exceeding 
a minimum nonzero value obtained from the first two terms in the limit 
of vanishing $\eta$ binding. Eq.~(\ref{eq:sqrt{s}}) in the limit $A\gg 1$ 
coincides with the nuclear-matter expression used in Refs.~\cite{FGM13,CFG14} 
for calculating $\eta$-nuclear quasibound states. 

\begin{figure}[htb] 
\begin{center} 
\includegraphics[width=1.0\textwidth]{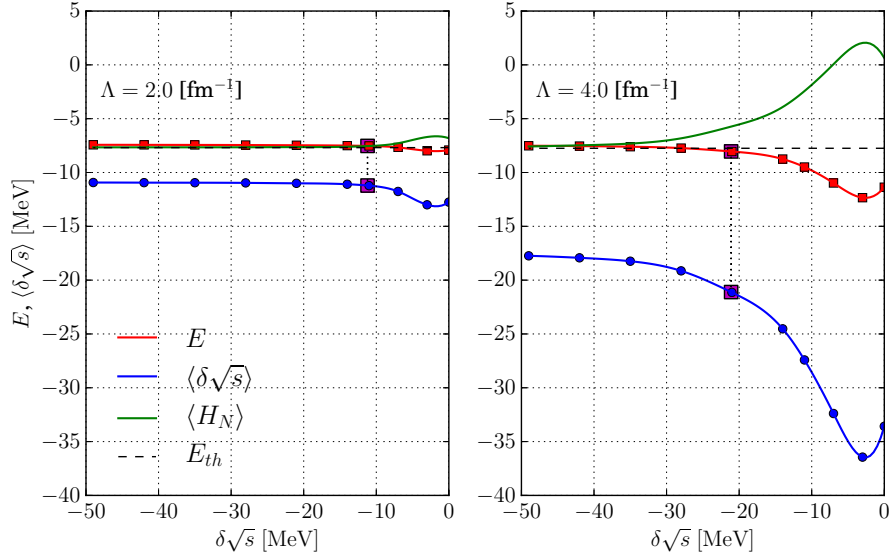} 
\caption{Calculated $\eta\, {^3{\rm He}}$ bound-state energies $E$ (squares) 
and expectation values $\langle\delta\sqrt{s}\rangle$ (circles) from 
Eq.~(\ref{eq:sqrt{s}}), using LO \nopieft $NN$, $NNN$ and $\eta NN$ regulated 
contact terms plus $\eta N$ potentials $v_{\eta N}^{\rm GW}(E)$ for two values 
of the scale parameter $\Lambda$, as a function of the input energy shift 
$\delta\sqrt{s}$ used for the energy argument of $v_{\eta N}^{\rm GW}(E)$. 
The dashed vertical line marks the self consistent values of $E$ and $\langle
\delta\sqrt{s}\rangle$. The dashed horizontal line marks the $^3$He core g.s. 
energy serving as threshold for a bound $\eta$, and the curve above it shows 
the squeezed core energy $\langle H_N \rangle$.} 
\label{fig:eta3He_EFT_GW} 
\end{center} 
\end{figure} 

Fig.~\ref{fig:eta3He_EFT_GW} demonstrates how the self consistency requirement 
works in actual calculations. The three curves plotted in each panel are 
obtained by interpolating a sequence of calculated $\eta\,^3$He bound-state 
energies (squares) and the corresponding expectation values $\langle\delta
\sqrt{s}\rangle$ (circles) from Eq.~(\ref{eq:sqrt{s}}) for $A=3$, as a 
function of the input $\delta\sqrt{s}$ to the energy argument $E_{\rm th}+
\delta\sqrt{s}$ of $v_{\eta N}^{\rm GW}$, for two choices of the momentum 
scale parameter $\Lambda=2,\,4$~fm$^{-1}$. The dashed vertical line marks the 
self consistent value of $\delta\sqrt{s}$ at which the outcome bound-state 
energy $E(\eta\,{^3{\rm He}})$ is evaluated, and the dashed horizontal line 
marks the $^3$He core energy $E(^3$He). Note that the self consistent value 
$E_{\rm sc}(\eta\,{^3{\rm He}})$ is {\it higher} than $E(^3$He) in the left 
panel for $\Lambda=2$~fm$^{-1}$, while it is {\it lower} than $E(^3$He) in 
the right panel for $\Lambda=4$~fm$^{-1}$. This means that, correspondingly, 
$\eta\,^3$He is slightly unbound for $\Lambda=2$~fm$^{-1}$ while slightly 
bound for $\Lambda=4$~fm$^{-1}$. We note furthermore that for threshold values 
$v_{\eta N}^{\rm GW}(E_{\rm th})$, i.e. $\delta\sqrt{s}=0$, $\eta\,^3$He is 
bound in both cases (and also if the often used but unfortunately unfounded 
self consistency requirement \cite{Garcia02} $\delta\sqrt{s}=E_{\eta}$ is 
imposed). Finally, the upper curve in Fig.~\ref{fig:eta3He_EFT_GW} shows the 
expectation value $\langle H_N \rangle$ of the nuclear core energy, which is 
bounded from below by the $^3$He core energy $E(^3$He) marked by the dashed 
horizontal line.

\section{Results and Discussion} 
\label{sec:res} 

Separation energies $B_{\eta}\equiv B(\eta\,{^A{\rm He}})-B(^A$He) 
(often called $\eta$ binding energies) of the $\eta\,^A$He isotopes 
with $A=3,4$ were calculated self consistently in the SVM using LO 
\nopieft $NN$, $NNN$ and $\eta NN$ regulated contact terms introduced 
in Eqs.~(\ref{eq:ij})--(\ref{eq:rij}) and Footnote~2, and a regulated 
$\eta N$ energy dependent contact term specified by Eq.~(\ref{eq:v(E)}), with 
scale parameters $\Lambda=2,4,6,8$~fm$^{-1}$. Two $\eta N$ coupled-channels 
models were used, GW~\cite{GWy05} and CS~\cite{CSm13}. The CS $\eta N$ 
interaction was found by far too weak to bind $\eta\,^3$He, and only by 
a fraction of MeV short of binding $\eta\,^4$He. The binding energies 
$B_{\eta}$ were evaluated using real Hamiltonians in which Im~$v_{\eta N}$ 
was disregarded. Restoring Im~$v_{\eta N}$ in optical model calculations was 
found particularly important for near-threshold bound states, lowering their 
calculated $B_{\eta}$ by 0.2$\pm$0.1~MeV. The $\eta$-nuclear widths $\Gamma_{
\eta}$ were calculated with wavefunctions $\Psi_{\rm g.s.}$ generated by these 
real Hamiltonians:  
\begin{equation}
\Gamma_{\eta}=-2\,\langle\,\Psi_{\rm g.s.}\, | \, {\rm Im} \, V_{\eta} \, | \, 
\Psi_{\rm g.s.} \, \rangle \;.
\label{eq:Gamma} 
\end{equation} 
Here, $V_{\eta}$ sums over all pairwise $\eta N$ interactions. Since 
$|{\rm Im}\,V_{\eta}|\ll |{\rm Re}\,V_{\eta}|$, this is a reasonable 
approximation. 

\begin{table}[htb] 
\begin{center} 
\caption{$\eta$ binding energies and widths (MeV) in the He isotopes from SVM 
calculations using $\eta N$ potentials $v_{\eta N}^{\rm GW}(E)$ with scale 
parameters $\Lambda=2,4$~fm$^{-1}$, together with the corresponding self 
consistent values of the downward energy shift (in MeV) $\delta\sqrt{s_{
\rm sc}}$. The values of $\Gamma(\eta\,^3$He) shown here outdate the 
erroneous, too large widths listed in Ref.~\cite{BFG15}.} 
\begin{tabular}{ccccccc} 
\hline\hline 
 & \multicolumn{3}{c}{$\eta\,^3$He} & \multicolumn{3}{c}{$\eta\,^4$He} \\ 
$\Lambda$~(fm$^{-1})$ & $\delta\sqrt{s_{\rm sc}}$ & $B_{\eta}$ & 
$\Gamma_{\eta}$ & $\delta\sqrt{s_{\rm sc}}$ & $B_{\eta}$ & $\Gamma_{\eta}$ \\ 
\hline 
2 & $-$11.2 & $-$0.16 & 0.24 & $-$16.5 & $-$0.15 & 0.34 \\ 
4 & $-$21.1 & $+$0.30 & 1.46 & $-$32.2 & $+$1.54 & 2.82 \\ 
\hline\hline 
\end{tabular} 
\label{tab:Beta} 
\end{center} 
\end{table} 

Binding energies $B_{\eta}$ and widths $\Gamma_{\eta}$ resulting from 
these self consistent calculations are listed in Table~\ref{tab:Beta} for 
the $\eta N$ potentials $v_{\eta N}^{\rm GW}(E)$ with scale parameters 
$\Lambda=2,4$~fm$^{-1}$, values for which our self consistency procedure 
was demonstrated in Fig.~\ref{fig:eta3He_EFT_GW}. Higher values of $\Lambda$ 
exceed by far the momentum breakdown scale $q^{\rho}_{\rm high}$ introduced 
in our previous work~\cite{BFG15} for $N^{\ast}$(1535) resonance meson-baryon 
models in which the $\eta N$ scattering amplitude $F_{\eta N}$ is determined. 
Taken literally, this would mean that the GW $\eta N$ interaction hardly binds 
$\eta\,^3$He, if at all, and is likely to bind slightly $\eta\,^4$He, with 
$B_{\eta}$ of order 1~MeV. 

\begin{figure}[!ht] 
\begin{center} 
\includegraphics[width=0.65\textwidth]{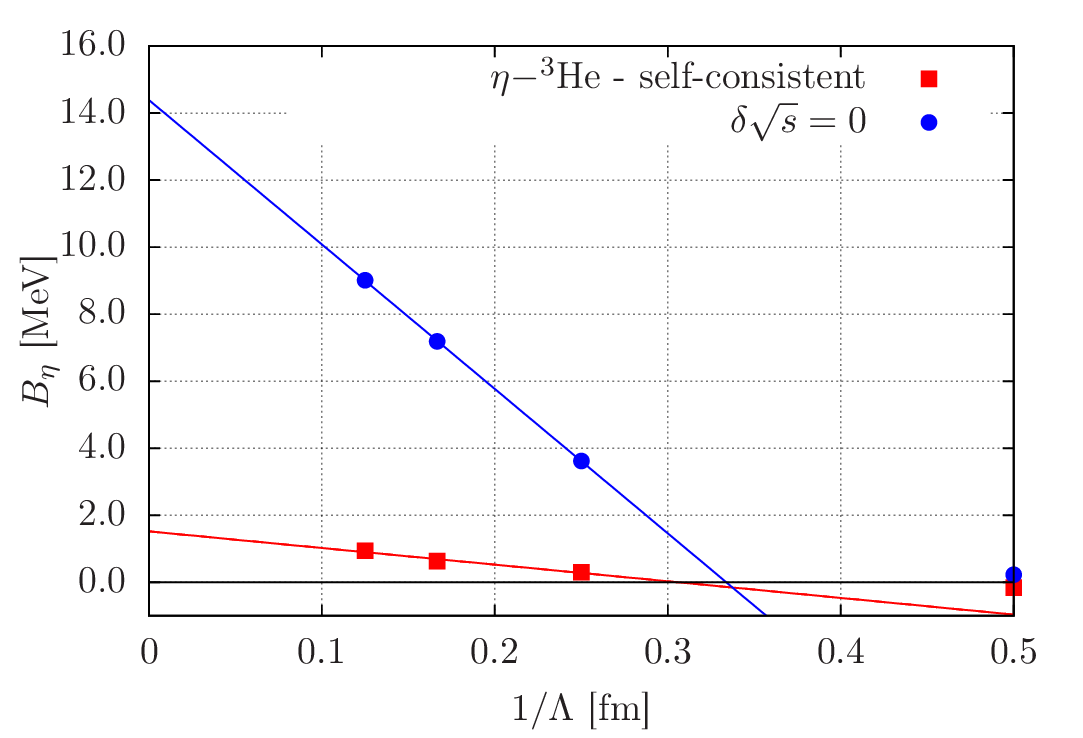}
\caption{$B_{\eta}(\eta\,{^3{\rm He}})$ as a function of $\Lambda^{-1}$, 
calculated using $\eta N$ potentials $v_{\eta N}^{\rm GW}(E)$ with scale 
parameters (from right to left) $\Lambda$=2,4,6,8~fm$^{-1}$. Squares 
(red) denote self consistent calculations, circles (blue) denote calculations 
with threshold values of the $\eta N$ interaction. Linear extrapolation to 
a point-like interaction, $\Lambda\to\infty$, is marked by straight lines.} 
\label{fig:3He_EFT} 
\end{center} 
\end{figure} 

\begin{figure}[!ht] 
\begin{center} 
\includegraphics[width=0.65\textwidth]{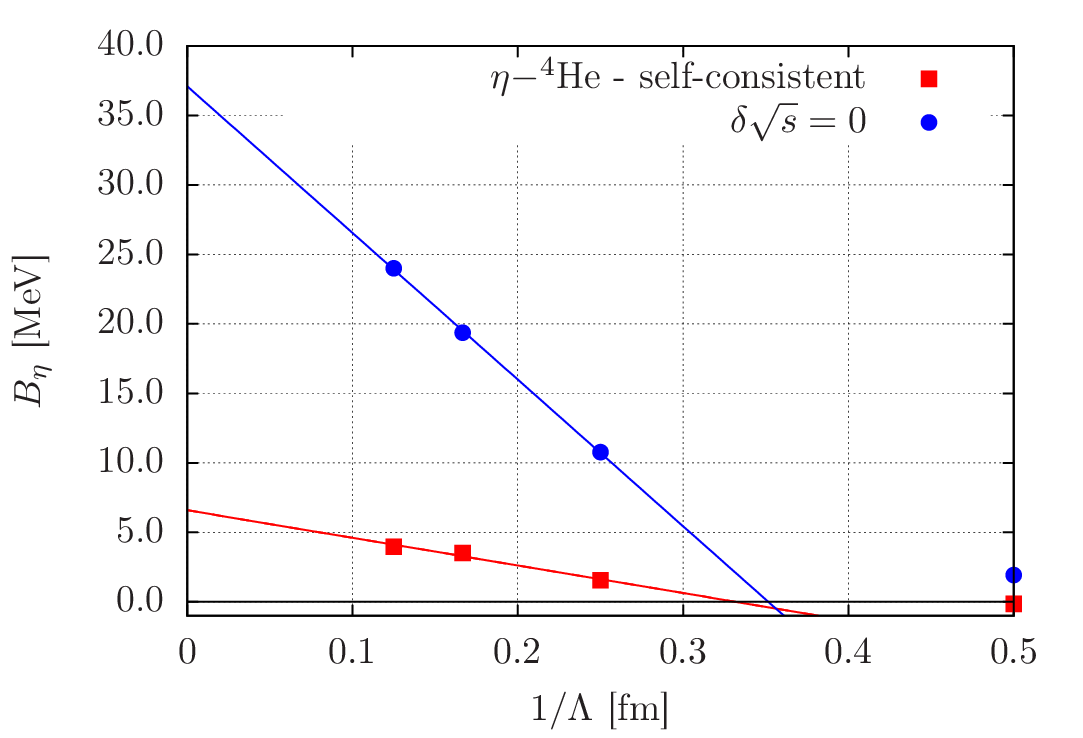} 
\caption{Same as in Fig.~\ref{fig:3He_EFT}, but for $\eta\,^4$He instead of 
$\eta\,^3$He.}  
\label{fig:4He_EFT} 
\end{center} 
\end{figure} 

Sequences of calculated values of $\eta$ binding energy $B_{\eta}$ using 
$\eta N$ potentials $v_{\eta N}^{\rm GW}(E)$ are shown for $\eta\,^3$He and 
$\eta\,^4$He in Figs.~\ref{fig:3He_EFT} and~\ref{fig:4He_EFT}, respectively, 
as a function of $\Lambda^{-1}$. The figures demonstrate that the larger 
$\Lambda$, the larger is the resulting $\eta$ binding energy $B_{\eta}$ 
in spite of a similar increase in the value of $-\delta\sqrt{s_{\rm sc}}$ 
in self consistent calculations which implies a weaker $\eta N$ potential 
strength $b(E_{\rm sc})$. The dependence of $B_{\eta}$ on $\Lambda$ is 
weak for $\eta\,^3$He and moderate for $\eta\,^4$He in these calculations. 
For $\Lambda\geq 4$~fm$^{-1}$, $B_{\eta}$ varies linearly in $\Lambda^{-1}$, 
with an average error of 50~keV for $\eta\,^3$He and 300~keV for $\eta\,^4$He, 
and with twice these errors upon extrapolating $\Lambda\to\infty$. 
In contrast, for calculations done at the $\eta N$ threshold, i.e. 
$\delta\sqrt{s}$=0, the resulting values of $B_{\eta}$ shown in the 
figures depend strongly on $\Lambda$ with almost perfect linear 
dependence on $\Lambda^{-1}$ for $\Lambda\geq 4$~fm$^{-1}$. 

Interestingly, Figs.~\ref{fig:3He_EFT} and~\ref{fig:4He_EFT} also suggest 
that $B_{\eta}$ assumes a {\it finite} value $B_{\eta}^{\Lambda\to\infty}$ 
in the limit of point $\eta N$ interaction. This follows directly from the 
introduction of a stabilizing $\eta NN$ LEC $d_{\eta NN}^{\Lambda}$ analogous 
to the $NNN$ LEC $d_{NNN}^{\Lambda}$ of Eq.~(\ref{eq:ijk}). Here we used 
a value $d_{\eta NN}^{\Lambda}=d_{NNN}^{\Lambda}$. The sensitivity of our 
results to this choice is studied in Appendix A. Generally, a Thomas 
collapse of three-body systems is averted in \nopieft by promoting 
a non-derivative three-body contact term from N$^2$LO to LO.

\section{Conclusion} 
\label{sec:concl} 

To summarize, we have presented genuine few-body SVM calculations of $\eta 
NNN$ ($\eta\,^3$He) bound states and, for the first time, also $\eta NNNN$ 
($\eta\,^4$He) bound states, using LO \nopieft interactions where the $\eta 
N$ interaction contact term was derived in coupled channels studies of the 
$N^{\ast}(1535)$ nucleon resonance. Special care was taken of the energy 
dependence of the input $\eta N$ subthreshold scattering amplitude by using 
a self consistency procedure. The present results exhibit renormalization 
scale invariance of the calculated $\eta$ binding energies owing to the 
introduction of a repulsive $\eta NN$ contact interaction. For physically 
motivated values of $\Lambda$, the onset of $\eta\,^3$He binding occurs for 
Re$\,a_{\eta N}$ close to 1~fm, as in model GW~\cite{GWy05}, consistently 
with our previous hyperspherical-basis $\eta NNN$ calculations~\cite{BFG15}. 
The onset of $\eta\,^4$He binding requires a lower value of Re~$a_{\eta N}$, 
exceeding however 0.7~fm; it is comfortably satisfied in model GW but not in 
model CS~\cite{CSm13}. Further dedicated experimental searches for $\eta\,^4
$He bound states are desirable in order to confirm the recent negative report 
from WASA-at-COSY~\cite{AAB16} which, taken at face value, implies that 
Re~$a_{\eta N}\lesssim 0.7$~fm. Similar results and conclusions hold valid 
in SVM calculations using non-EFT realistic nuclear models~\cite{MNC77,AV402} 
augmented by the same $\eta N$ interaction models used here, and will be 
reported elsewhere.

\section*{Acknowledgments} 

We thank Ji\v{r}\'{i} Mare\v{s} and Martin Schaefer for useful discussions 
on related matters. This work was supported in part (NB) by the Israel 
Science Foundation grant 1308/16, in part (NB, BB) by Pazi Fund grants, 
and in part (EF, AG) by the EU initiative FP7, Hadron-Physics3, under the 
SPHERE and LEANNIS cooperation programs.

\section*{Appendix A: Erratum \cite{BBFG17E} to ``Onset of $\eta$-nuclear 
binding in a pionless EFT approach" \cite{BBFG17}} 

A three-body $\eta NN$ force was inadvertently overlooked in the 
potential model description and discussion in Ref.~\cite{BBFG17}. 
In the actual calculations, however, the LO interaction between the 
$\eta$ and the nucleons was composed of the $\eta N$ term discussed 
here in Sect.~2.3, supplemented by an $\eta NN$ term 
\begin{equation} 
V_{\eta N_i N_j} = d_{\eta N N}^\Lambda \delta_{\Lambda}(r_{\eta N_i},
r_{\eta N_j}). 
\label{eq:etaNN} 
\end{equation} 
In this expression, $\delta_{\Lambda}(r_{\eta N_i},r_{\eta N_j})$ is a 
product of normalized pairwise Gaussians $\delta_{\Lambda}(r_{\eta N_i})$ and 
$\delta_{\Lambda}(r_{\eta N_j})$, with range parameter inversely proportional 
to the momentum-scale parameter $\Lambda$, as defined by Eq.~(\ref{eq:rij}) 
here. For the results presented in this paper, the low energy constant (LEC) 
$d_{\eta N N}^{\Lambda}$ was set equal to the nuclear $NNN$ LEC $d_{NNN}^{
\Lambda}$. Setting $d_{\eta N N}^{\Lambda}=0$, the $\eta$-deuteron ($\eta\,d$) 
system, and therefore any $\eta$-nucleus system, would collapse as $\Lambda
\rightarrow\infty$. 

\begin{figure}[!ht] 
\begin{center} 
\includegraphics[width=0.65\textwidth]{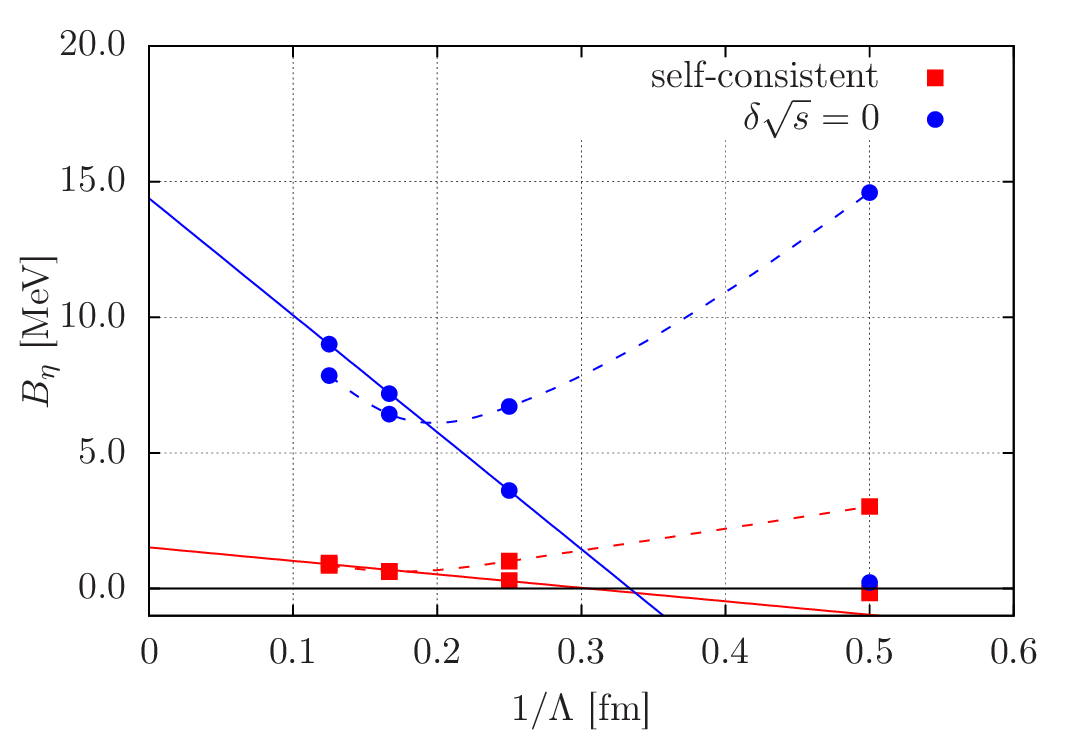} 
\caption{$B_{\eta}(\eta\,^3$He) as a function of $1/\Lambda$, calculated using 
$\eta N$ potentials $v_{\eta N}^{\rm GW}(E)$ for two choices of the $\eta NN$ 
LEC. Solid lines: $d^{\Lambda}_{\eta NN}=d^{\Lambda}_{NNN}$~\cite{BBFG17}, 
dashed lines: $d^{\Lambda}_{\eta NN}$ fitted to produce $B_{\eta}(\eta\,d)=0$. 
Self consistent calculations are marked by squares (red); calculations using 
threshold values $v_{\eta N}^{\rm GW}(E_{\rm th})$ are marked by spheres 
(blue). Linear extrapolations to a point-like interaction, $\Lambda\to\infty$, 
are marked by straight lines.} 
\label{fig:3He_EFTrev} 
\end{center} 
\end{figure}

\begin{figure}[!ht] 
\begin{center} 
\includegraphics[width=0.65\textwidth]{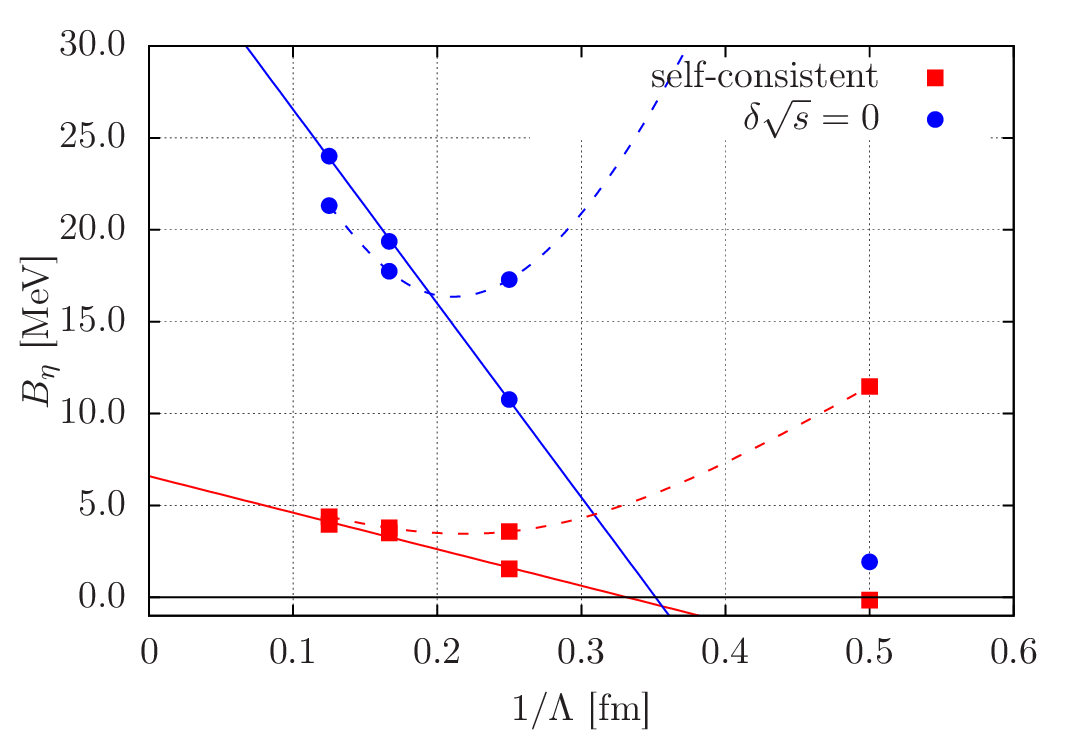} 
\caption{Same as in Fig.~\ref{fig:3He_EFTrev}, but for $\eta\,^4$He instead 
of $\eta\,^3$He.} 
\label{fig:4He_EFTrev} 
\end{center} 
\end{figure}

The parameter $d^{\Lambda}_{\eta N N}$ is a free parameter to be fixed by 
experimental data. In the absence of such data one may estimate its value 
using the nuclear $NNN$ LEC, $d^{\Lambda}_{\eta N N}=d^{\Lambda}_{NNN}$, 
as done here~\cite{BBFG17}, or to set a bound on its value accepting 
that $\eta\,d$ is unbound~\cite{KWi15}, i.e. set $d^{\Lambda}_{\eta N N}$ 
so that $B_{\eta}(\eta\,d)=0$. To check the sensitivity of the present 
results~\cite{BBFG17} to these distinct choices of $d^{\Lambda}_{\eta N N}$, 
we show in Figs.~\ref{fig:3He_EFTrev} and~\ref{fig:4He_EFTrev} $\eta$ 
separation energies $B_{\eta}$ in $\eta\,^3$He and $\eta\,^4$He, respectively, 
calculated using $\eta N$ potentials $v_{\eta N}^{\rm GW}(E)$ under these 
two choices of $d^{\Lambda}_{\eta N N}$. Figs.~\ref{fig:3He_EFTrev} 
and~\ref{fig:4He_EFTrev} update the original Figs.~\ref{fig:3He_EFT} 
and~\ref{fig:4He_EFT} here~\cite{BBFG17}. 

Figures \ref{fig:3He_EFTrev} and~\ref{fig:4He_EFTrev} demonstrate that the 
two choices made for the three-body $\eta NN$ LEC yield practically identical 
values of $B_{\eta}$ in the limit $\Lambda\rightarrow\infty$. For values 
f $\Lambda$ near the physical breakdown scale $\Lambda\approx 4$~fm$^{-1}$, 
however, $B_{\eta}$ differs by about 0.7~MeV for $\eta\,^3$He and 2~MeV for 
$\eta\,^4$He between the two choices applied in self consistent calculations 
(lower group of curves). Since $\eta\,d$ is unbound~\cite{KWi15}, the choice 
marked in dashed lines in both figures is likely to somewhat overestimate 
$B_{\eta}$. Nevertheless, these $\eta$ separation energies are in good 
agreement with the non-EFT $B_{\eta}$ values calculated recently using 
the same two-body energy dependent $\eta N$ interaction~\cite{BFG17}.

\end{document}